\documentclass[conference]{IEEEtran}
\usepackage[a4paper, total={184mm,239mm}]{geometry}
\def\BibTeX{{\rm B\kern-.05em{\sc i\kern-.025em b}\kern-.08em
    T\kern-.1667em\lower.7ex\hbox{E}\kern-.125emX}}
\usepackage{bm}
\usepackage{epsfig}
\usepackage{graphicx}
\usepackage{subfigure}
\usepackage{float}
\usepackage{cite}
\usepackage{url}
\usepackage{xcolor}
\usepackage{balance}
\usepackage{mdwlist}
\usepackage{multirow}
\usepackage{threeparttable}
\usepackage{enumitem}
\usepackage{amsmath}
\usepackage{stmaryrd}
\usepackage{booktabs}
\usepackage{siunitx}
\usepackage{amsthm}
\usepackage{mathdots}
\usepackage{threeparttable}
\usepackage{epsfig,amsfonts}
\usepackage{amsfonts}
\usepackage{adjustbox}
\usepackage{mathtools}

\usepackage{algpseudocode}
\usepackage{algorithm}

\def\R{\mathbb{R}}
\def\E{\mathbb{E}}

\newcommand{\argmin}{\mathop{\mathrm{argmin}}}

\newcommand{\st}{\mathop{\mathrm{s.t.}}}

\newcommand{\parm}{{\xi}}
\newcommand{\vecpar}{\boldsymbol{\parm}}

\newcommand{\mat}[1]{\mathbf{#1}}
\newcommand{\vect}[1]{\boldsymbol{#1}}

\begin{document}
\title{Distributionally Robust Circuit Design Optimization under Variation Shifts}

\author{
Yifan Pan$^1$$^*$, Zichang He$^1$$^*$, Nanlin Guo$^2$ and Zheng Zhang$^1$\\
\IEEEauthorblockA{$^1$Department of Electrical and Computer Engineering, 
University of California, Santa Barbara, CA 93106\\
$^2$School of Microelectronics, Fudan University, Shanghai, China
}
Emails: yifanpan@ucsb.edu, zichanghe@ucsb.edu, nlguo21@m.fudan.edu.cn, zhengzhang@ece.ucsb.edu\\
$^*$ Both authors contributed equally
}
\maketitle
\begin{abstract}
Due to the significant process variations, designers have to optimize the statistical performance distribution of nano-scale IC design in most cases. This problem has been investigated for decades under the formulation of stochastic optimization, which minimizes the expected value of a performance metric while assuming that the distribution of process variation is exactly given. This paper rethinks the variation-aware circuit design optimization from a new perspective. First, we discuss the variation shift problem, which means that the actual density function of process variations almost always differs from the given model and is often unknown. Consequently, we propose to formulate the variation-aware circuit design optimization as a distributionally robust optimization problem, which does not require the exact distribution of process variations. By selecting an appropriate uncertainty set for the probability density function of process variations, we solve the shift-aware circuit optimization problem using distributionally robust Bayesian optimization. This method is validated with both a photonic IC and an electronics IC. Our optimized circuits show excellent robustness against variation shifts: the optimized circuit has excellent performance under many possible distributions of process variations that differ from the given statistical model. This work has the potential to enable a new research direction and inspire subsequent research at different levels of the EDA flow under the setting of variation shift. %
\end{abstract}
\begin{IEEEkeywords}
Process variation; variation shift; distributionally robust optimization; distributionally robust Bayesian optimization
\end{IEEEkeywords}

\section{Introduction}
\label{introduction}

In semiconductor chip design, imperfect nano-fabrications have led to dramatic performance degradation and yield loss. These variations become even more pronounced in emerging computing technologies. The EDA (electronic design automation) community has long been engaged in exploring variation-aware simulation~\cite{singhee2010quasi,stievano2011parameters,zhang2013stochastic,zhang2014enabling,he2021high}, modeling~\cite{QBPV,ReviewSM,zhang2016big}, and optimization techniques~\cite{srivastava2005statistical,liu2011efficient,wang2017efficient} for the design and fabrication of integrated circuits, MEMS and photonics.

In most existing approaches, it is assumed that the process variations are described {\it exactly} by a probability density function (PDF). Under this assumption, variation-aware circuit optimization~\cite{srivastava2005statistical,liu2011efficient,wang2017efficient,cui2020chance,he2021pobo,zhang2020bayesian} has been formulated as a stochastic optimization problem: the goal is to minimize the expectation value of a cost function subject to some deterministic or stochastic design constraints. Compared to robust optimization-based design~\cite{antreich1994circuit,dharchoudhury1995worst} that optimizes worst-case circuit performance, stochastic optimization leads to more accurate and less conservative results by accounting for the PDF of process variations. So far, stochastic optimization-based approaches have achieved great success in the EDA field.

In this paper, we ask a fundamental question: {\it what if the PDF of process variations is uncertain and/or not exactly known?} This is a critical yet rarely explored question in EDA. In practical EDA flow, the PDF of process variations is normally extracted based on some measurement data of a foundry. As will be explained in Section~\ref{sec:variation_shift}, the given PDF often differs from the actual one due to the limited size and low quality of measurement data, as well as due to unavoidable errors in the statistical modeling process. Even if the given PDF is accurate enough at the beginning, the actual PDF may still change significantly over time. We call the phenomena of the given PDF differing from the actual one as {\it variation shifts}. In this paper, we investigate how to optimize a circuit design under such variation shifts. Definitely, worst-case circuit optimization techniques~\cite{antreich1994circuit,dharchoudhury1995worst} can still be applied for some cases when variation shift exists and the variation parameters are bounded, at the cost of (probably tremendously) over-conservative design. However, our goal is to develop a rigorous problem formulation and a proof-of-concept solver for shift-aware circuit optimization by considering both the statistical nature and the unknown PDF of process variations.

\begin{figure*}[t]
	\centering
	\includegraphics[width=6.8in]{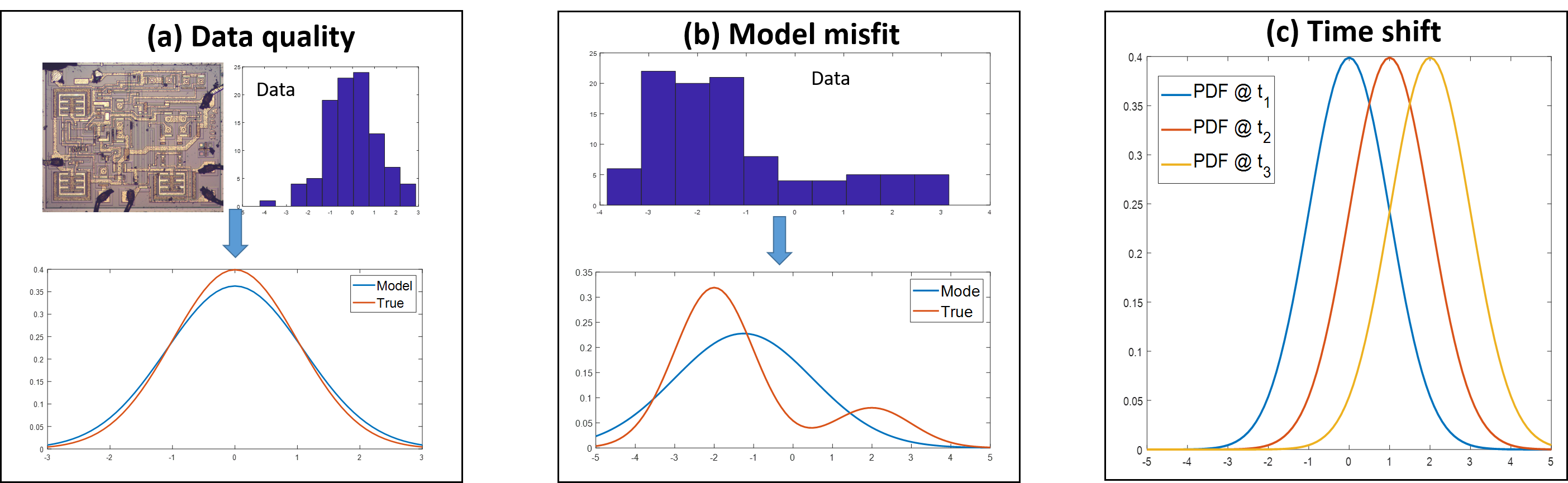}
\caption{Various sources of variation shifts. (a): Insufficient and/or inaccurate measurement data can lead to inaccurate density estimation of process variations. (b): An imperfect chosen (e.g., over-simplified) model can generate a distribution model that is far away from the true one. (3) The distribution of device parameters can shift over time. In all cases, stochastic optimization can produce underperforming results when one uses a fixed PDF model (e.g., a nominal PDF) that differs from the unknown actual one. 
} 
\label{fig:distribution_shift}
\end{figure*}

\textbf{Paper contributions.}
In this work, we investigate the problem formulation, numerical solver, and validation of shift-aware circuit optimization. Our specific contributions include:
\begin{itemize}[leftmargin=*]
    \item We present, for the first time, a mathematical formulation for the shift-aware circuit optimization problem. Starting with an introduction of the typical sources of variation shifts in IC design, we formulate shift-aware circuit optimization as a distributionally robust optimization problem.
    \item We present numerical methods to solve the shift-aware optimization problem by extending the traditional variation-aware Bayesian circuit optimization methods~\cite{wang2017efficient,zhang2020bayesian}. Specifically, we employ an uncertainty ball whose radius is defined by $\varphi$-divergence to represent the distribution shift of process variations. By leveraging recent advancements of distributionally robust Bayesian optimization (DRBO)~\cite{kirschner2020distributionally,nguyen2020distributionally,husain2022distributionally}, we can solve our shift-aware circuit optimization problem effectively and efficiently.
    \item We validate our approach on both photonic and electronic IC benchmarks. Through numerical experiments on two realistic design cases, we demonstrate that the proposed distributionally robust optimization method can maintain excellent performance metrics and high yield under various unforeseen potential PDFs of process variations.
\end{itemize}
We regard this work as a first attempt to address the critical issue of variation shifts. This research could enable a new direction, and motivate the study of numerous shift-aware EDA problems in the future.

\section{Problem Formulation}
\subsection{Variation Shifts}
\label{sec:variation_shift}
In classical statistical variation-aware circuit design, process variations are assumed to be described precisely by a reference probability density function (PDF) $\rho_0(\vecpar)$ (named \emph{nominal distribution} in this paper). Consequently, a circuit optimization framework seeks to minimize the expected value of a cost function $f(\mat{x},\vecpar)$ over a design variable $\mat{x}$ within its design space $\cal X$. This is mathematically expressed as: 
\begin{equation}
\label{eq:standard_opt}
 \min \limits_{\mat{x} \in {\cal X}} \mathbb{E}_{\rho_0(\vecpar)}\left[ f(\mat{x},\vecpar) \right].   
\end{equation}

A typical variation-aware design workflow consists of three steps. (1) Testing engineers measure some data samples $\{ \vecpar ^i\}_{i=1}^M$ of device parameters from specially designed testing circuits. (2) A nominal PDF $\rho_0(\vecpar)$ is extracted from the testing data to describe the process variations. (3) The extracted PDF $\rho_0(\vecpar)$ is then used in an EDA tool to conduct statistical circuit simulation, modeling, and optimization. 
However, the true PDF of process variations, denoted as $\rho(\vecpar)$, is rarely identical to $\rho_0(\vecpar)$ due to the following reasons. 

{\bf Poor data quality.} Ideally, we can approximate the true PDF $\rho(\vecpar)$ with arbitrarily high accuracy if we have an infinite number of i.i.d. data samples $\{ \vecpar^i\}$ that precisely follow the distribution $\rho(\vecpar)$. However, in practice, this ideal situation rarely happens since the measurement data samples could be very noisy~\cite{stine1997analysis}. Additionally, since fabricating and measuring testing circuits can be costly, we typically only have access to a limited number of data samples, further restricting the modeling accuracy of $\rho_0(\vecpar)$.

{\bf Model misfit.} In reality, process variations often have a complicated (joint) PDF $\rho(\vecpar)$ (e.g., Gaussian mixture distribution) due to the multi-modal behavior and correlation among random parameters~\cite{stine1997analysis,gmm15}. However, in engineering practice, $\rho_0(\vecpar)$ is often chosen as a trivial distribution (e.g., correlated or independent normal distribution) for simplicity~\cite{yu2013statistical,raj2011nanoscale}. This inherently introduces a discrepancy between the true and modeled distributions.
  
{\bf Time shift.} It is well known that device parameters can shift over time~\cite{reddy2004impact,han2011, wang2007efficient} due to many factors, such as reliability issues and external environmental impact (e.g., radiation, temperature fluctuations). This phenomenon (including but not limited to the aging effects) implies that even if the initially modeled $\rho_0(\vecpar)$ accurately describes the process variations, its accuracy may decrease as time evolves.

Fig.~\ref{fig:distribution_shift} visualizes the mismatch between the modeled PDF $\rho_0(\vecpar)$ and true PDF $\rho(\vecpar)$ under the above three scenarios.
\begin{figure}[t]
    \centering
    \includegraphics[width = 3.3in]{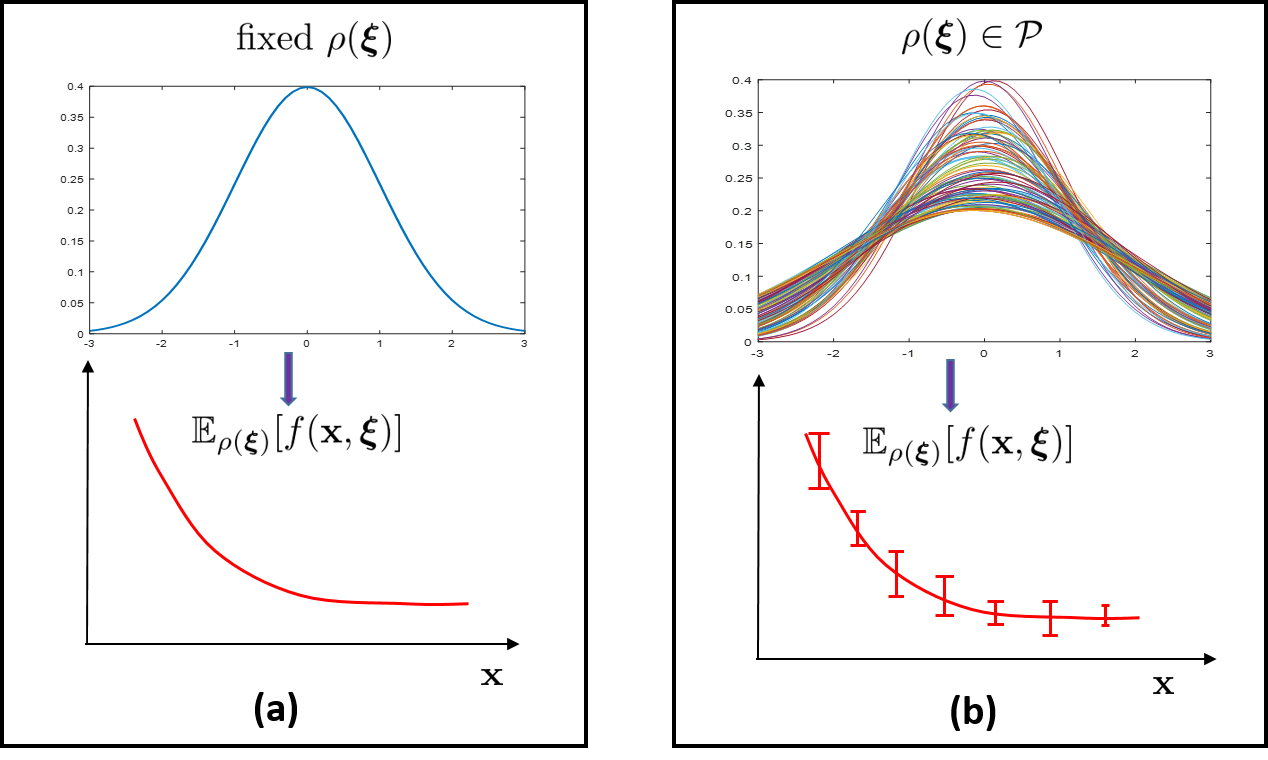}
    \caption{Comparison between traditional stochastic circuit optimization and our proposed optimization. (a) In classical variation-aware optimization, $\rho(\vecpar)$ is assumed to be fixed and known, allowing for a unique determination of the cost function $\mathbb{E}_{\rho(\vecpar)}\left[f(\mat{x},\vecpar) \right]$ for any given design variable $\mat{x}$.
    (b) Under the presence of process variation shifts, the true density $\rho(\vecpar)$ becomes unknown. This consequently introduces uncertainty into the cost function $\mathbb{E}_{\rho(\vecpar)}\left[ f(\mat{x},\vecpar) \right]$, complicating the optimization process.
    }
    \label{fig:DROpt_compare}
\end{figure}

\subsection{Formulation via Distributionally Robust Optimization}
Due to the variation shifts, we need to rethink the fundamental problem formulation of variation-aware circuit optimization. As shown in Fig.~\ref{fig:DROpt_compare} (a), in conventional problem settings, given a particular $\mat{x}$, the cost function $\mathbb{E}_{\rho(\vecpar)}\left[ f(\mat{x},\vecpar) \right]$ can be uniquely determined, allowing the search for its minimum value. However, the exact $\rho(\vecpar)$ is unknown due to variation shifts, which introduces uncertainties to the cost function $\mathbb{E}_{\rho(\vecpar)}\left[ f(\mat{x},\vecpar) \right]$ as shown in Fig.~\ref{fig:DROpt_compare} (b). Even given a specific $\mat{x}$, we cannot determine the exact value of this cost function. Therefore, we cannot simply minimize $\mathbb{E}_{\rho(\vecpar)}\left[ f(\mat{x},\vecpar) \right]$. 

To address the above challenge, we propose a novel formulation for a generic variation-aware circuit optimization under variation shifts. Specifically, we describe the design optimization as the following {\it distributionally robust optimization}: 
\begin{equation}\label{eq:shift_aware_minmax}
    \min_{\mat{x} \in {\cal X}} \sup_{\rho(\vecpar) \in {\cal P}} \quad \mathbb{E}_{\rho(\vecpar)}[f(\mat{x},\vect{\xi})],
\end{equation}
where ${\cal P}$ is an uncertainty set that includes all possible PDFs of the variation parameters $\vect{\xi}$. Intuitively, we will minimize the upper bound of $\mathbb{E}_{\rho(\vecpar)}[f(\mat{x},\vect{\xi})]$ by considering the variation shifts. In the extreme case where ${\cal P}$=$\{ \rho_0(\vecpar)\}$, our formulation~\eqref{eq:shift_aware_minmax} simplifies to a standard stochastic optimization \eqref{eq:standard_opt}.

One important case is yield-aware optimization:
\begin{equation}\label{eq:yield_constrained}
\begin{aligned}
\min_{\mat{x} \in {\cal X}} \quad& \E_{\rho(\vecpar)} [f_{\rm obj}(\mat{x},\vect{\xi})]\\
\st \quad & {\rm Prob}\left [g_k\left (\mat{x}, \vect{\xi}\right) \leq  0, \forall k\in [n] \right]\geq 1-\tau,   
\end{aligned}
\end{equation}
where $f_{\rm obj}$ is a design objective,  $g_i(\mat{x}, \vect{\xi})>0$ describes a violation of the design specification, and $\tau$ is a risk tolerance. This formulation aims to optimize the design objective while satisfying a yield requirement~\cite{cui2020chance,he2021pobo}. To handle the yield constraints, we incorporate these constraints as a penalty term into the objective function: 
\begin{equation}\label{eq:penalized_cost}
\begin{aligned}
        f(\mat{x},\vect{\xi}) = {f_{\rm obj}}(\mat{x},\vect{\xi})+ \lambda I(\mat{x},\vect{\xi}) ~ \text{with}\\
    I(\mat{x},\vect{\xi}) = 
    \begin{cases}
    0, & g_k(\mat{x},\vect{\xi})\le 0, \forall k \in [n],\\
    1, & \text{otherwise}.
    \end{cases}
\end{aligned}
\end{equation}
Here, $I(\mat{x},\vect{\xi})$ is an indicator function for risk violations and $\lambda \ge 0$ is a parameter for penalizing the constraint violations. The yield $1-\tau$ is defined as the probability that all risk constraints are met, i.e., ${1-\E_{\rho(\xi)}[I(\mat{x},\vect{\xi})]}$.
The resulting cost function $f(\mat{x},\vect{\xi})$ can then be integrated in the shift-aware optimization defined in~\eqref{eq:shift_aware_minmax}. %

\section{Distributionally Robust Bayesian Optimization Solver}
The distributionally robust circuit optimization as defined in~\eqref{eq:shift_aware_minmax} may be intractable in practice because: (a) the uncertainty set ${\cal P}$ may contain an infinite number of PDFs describing process variations; (b) the min-max problem is inherently hard to solve; (c) we do not have an analytical form for $f(\mat{x},\vect{\xi})$ and its simulation could be costly. 
To tackle these challenges, we will first define the PDF uncertainty set ${\cal P}$ appropriately. 
Subsequently, we utilize distributionally robust Bayesian optimization (DRBO)~\cite{kirschner2020distributionally,nguyen2020distributionally,husain2022distributionally}, a recently developed technique in the machine learning community, to efficiently solve problem~\eqref{eq:shift_aware_minmax}.

\subsection{Distribution Uncertainty Set}
In our approach, we model the PDF uncertainty set ${\cal P}$ as a ball whose center is the nominal distribution ${\rho_0}(\vecpar)$ (which is often an inaccurate PDF provided by a foundry) and whose radius $\varepsilon$ is measured by a distribution divergence $\mathcal{D}$:
\begin{equation}\label{eq:def_ball}
    {\cal P}:= \mathcal{B}(\rho_0) =\{\rho: \mathcal{D}({\rho_0}, \rho) \leq \varepsilon \}.
\end{equation}
Here $\mathcal{D}({\rho_0}, \rho)$ measures the difference between $\rho_0(\vecpar)$ and $\rho(\vecpar)$. In practice, we can decide the value of radius $\epsilon$ based on estimations of variation shifts.  

Among several choices for the divergence $\mathcal{D}$~\cite{rahimian2019distributionally}, we choose the $\varphi$-divergence (also called as $f$-divergence) for its advantages in computational efficiency. Let $\rho$ and $\rho_0$ be two distributions such that $\rho$ is absolutely continuous with respect to $\rho_0$, the $\varphi$-divergence from $\rho$ to $\rho_0$ is defined as %
\cite{f_diver_1966}:
\begin{equation}\label{eq:divergence_define}
\mathcal{D}_{\varphi} (\rho, \rho_0) \overset{\Delta}{=}\mathbb{E}_{\rho_0}\left[\varphi \left(\dfrac{d\rho}{d\rho_0}(\vect{\xi})\right)\right],
\end{equation}
where $\varphi: \R \rightarrow (-\infty,\infty]$ is a
convex, lower semi-continuous function such that is $\varphi(1) = 0$ and ${d\rho}/{d\rho_0}$ is a \emph{Radon-Nikodym} derivative. %
In this paper, we choose $\varphi(u)=(u-1)^2$ to measure the radius of ${\cal P}$ by $\chi^2$-divergence. According to~\cite{husain2022distributionally}, we can simplify Problem~\eqref{eq:shift_aware_minmax} to a single-level optimization problem:
\begin{equation}\label{eq:chi2_minmax}
    \min_{\mat{x} \in \mat{X}} ~ \mathbb{E}_{\rho_0(\vect{\xi})}[f(\mat{x},\vect{\xi})] + \sqrt{\varepsilon \cdot \text{Var}_{\rho_0(\vect{\xi})}[f(\mat{x},\vect{\xi})]}.
\end{equation}

\subsection{DRBO Workflow} 

Next, we explain how to solve~\eqref{eq:chi2_minmax} via DRBO with a few circuit simulation samples. Similar to a standard Bayesian optimization (BO), DRBO sequentially builds a probabilistic surrogate model of $f(\mat{x},\vect{\xi})$ and explores the design space by minimizing an acquisition function. The algorithm is summarized in Algorithm~\ref{alg:DRBO_alg}, and the key steps are explained below.
\begin{itemize}[leftmargin=*]
    \item {\bf Step 1}. Build a probabilistic surrogate model of $f(\mat{x},\vect{\xi})$. Here we choose a Gaussian process model, which has been widely used in BO-based circuit optimization. 
    \item {\bf Step 2}. Minimize the acquisition function $A(\mat{x})$: 
    \begin{equation}\label{eq:acqusition_func}
        \mat{x} = \argmin_{\mat{x}\in \mat{X}} A(\mat{x}).
    \end{equation}
    We use the lower confidence bound (LCB) to define $A(\mat{x})$. This approach allows us to effectively balance between exploiting promising current designs and exploring the design space characterized by significant model uncertainty:
    \begin{equation}\label{eq:chi_square_acq}
    \begin{split}
        A(\mat{x}) :=  \frac{1}{L}  \sum_{l=1}^L[\mu(\mat{x},\vect{\xi}^l)-\sqrt{\beta}\sigma(\mat{x},\vect{\xi}^l)] + \\
        \sqrt{\frac{\varepsilon}{L}\sum_{l=1}^L{(\mu(\mat{x},\vect{\xi}^l)-\bar{\mu})}^2},
    \end{split}
    \end{equation}
    where $\mu(\mat{x},\vect{\xi})$ and $\sigma(\mat{x},\vect{\xi})$ represent the predictive mean and standard deviation offered by the probabilistic surrogate model, and $\beta$ is a balancing factor. A finite number of samples $\{\vect{\xi}^l\}_{l=1}^L$ sampled from $\rho_0$ are used to estimate the performance mean and variance over the nominal variation and $\bar{\mu}=\frac{1}{L}\sum_{l=1}^L {\mu(\mat{x},\vect{\xi}^l)}$ denotes the performance mean.
    \item {\bf Step 3}. If convergence is not achieved, we draw a new sample from $\rho_0$ and augment the existing sample set. Then the algorithm returns to Step 1.
\end{itemize}

\begin{algorithm}[t]
\caption{Overall DRBO algorithm}\label{alg:DRBO_alg}
\begin{algorithmic}[1]
\Require Initial sample set ${\mathcal{S}_{0}=\{\mat{x}^i,\vect{\xi}^i,f(\mat{x}^i,\vect{\xi}^i)\}_{i=1}^M}$, nominal PDF of variations $\rho_0$, uncertainty ball radius $\varepsilon$, maximum iteration $T$
\Ensure The optimal circuit design $\mat{x}^\star$ for Problem~\eqref{eq:chi2_minmax}
\For{$t = 1,2,...,T$}
\State Construct a probabilistic surrogate model based on $\mathcal{S}_{t-1}$
\State Solve the next query point $\mat{x}_{t}$ via~\eqref{eq:acqusition_func} with acquisition function~\eqref{eq:chi_square_acq}
\State {Sample variation $\vecpar_t \sim \rho_0$ and simulate $f(\mat{x}_t, \vect{\xi}_t)$}
\State {Augment data set $\mathcal{S}_{t} \leftarrow {\mathcal{S}_{t-1} \cup (\mat{x}_t, \vect{\xi}_t, f(\mat{x}_t, \vect{\xi}_t)})$}
\EndFor 
\State {Return the optimal design $\mat{x}^\star$}
\end{algorithmic}
\end{algorithm}

{\bf Remarks.} Compared with standard BO, the DRBO algorithm aims to find a robust solution under shifted variations by penalizing an additional term associated with the variance. In other words, without the second term in~\eqref{eq:chi2_minmax} and consequently in~\eqref{eq:chi_square_acq}, DRBO would simply reduce to standard BO.

\section{Implementation Details}
In this section, we explain some implementation details of the DRBO algorithm.

\textbf{Probabilistic surrogate model.} We choose the commonly used Gaussian processes (GP) as the probabilistic surrogate model. For simplicity, we denote ${\vect{\theta} = (\mat{x},\vect{\xi})}$. 
Let ${\vect{\Theta} = \{\vect{\theta}^i\}_{i=1}^M}$ be a set of training samples, and let ${\mat{y}=\{f({\vect{\theta}}^i)\}_{i=1}^M}$ be their simulation outputs. 
Given a pre-specified prior mean $m(\vect{\theta})$ and a kernel function $k(\vect{\theta},\vect{\theta}^\prime)$, a GP model assumes that the output $\mat{y}$ follow a Gaussian distribution:
\begin{equation}
    {\rm Prob}(\mat{y}) \sim \mathcal{N}(\mat{y}|\mat{m}, \mat{K}),
\end{equation}
where $\mat{m} \in \R^M$ is a mean vector, with the $i$-th element being $m(\vect{\theta}^i)$ and $\mat{K} = \mat{k}(\vect{\Theta},\vect{\Theta}) \in \R^{M \times M}$ is a covariance matrix with the $(i,j)$-th element being $k(\vect{\theta}^i,\vect{\theta}^j)$. For a new data $\vect{\theta}^\prime$, we can predict its posterior mean and variance from the GP model as follows:
\begin{equation}
    \begin{cases}
        \mu({\vect{\theta}^\prime}) = {\mat{k}(\vect{\theta}^\prime, \vect{\Theta})}^T{\mat{K}}^{-1} \mat{y} \\
        \sigma^2({\vect{\theta}^\prime}) = k(\vect{\theta}^\prime,\vect{\theta}^\prime) - {\mat{k}(\vect{\theta}^\prime,\vect{\Theta})}^T{\mat{K}}^{-1}\mat{k}(\vect{\Theta}, \vect{\theta}^\prime),
    \end{cases}
\end{equation}

We set the prior mean as ${m(\vect{\theta})=0}$ and the kernel as a Matern function. The hyperparameters in the GP model are optimized by maximizing the log marginal likelihood. %

\textbf{Improved modeling of the cost function.} 
The penalty term in~\eqref{eq:penalized_cost} may become non-smooth, introducing additional challenges for Gaussian process modeling. To address this issue, we employ two distinct GP models to separately estimate circuit performance and feasibility, with the feasibility being estimated by a Gaussian process classifier~\cite{williams2006gaussian}. %

\textbf{Stop criteria.} Choosing the optimal $\mat{x}$ after multiple iterations of sampling is not a trivial task, since our cost function involves an expectation over the variations, specifically ${\mathbb{E}_{\rho_0(\vect{\xi})}[f(\mat{x},\vect{\xi})]}$. Estimating this expectation through simulating $f(\cdot)$ would be computationally expensive. To address this, we select $\mat{x}$ with the minimal value of~\eqref{eq:chi2_minmax} according to the posterior distribution of $\hat{f}(\mat{x},\vect{\xi})$. This is a common strategy in such contexts~\cite{cakmak2020bayesian,frazier2018tutorial}. An alternative strategy for early stop is to return a stable solution after several consecutive iterations.

\textbf{Minimizing the acquisition function.} Many numerical optimization algorithms can be employed to minimize the acquisition function~\eqref{eq:acqusition_func}.
We simply adopt another Bayesian optimization as the optimizer.

In Line 4 of Algorithm~\ref{alg:DRBO_alg}, only one variation sample is drawn from the nominal PDF $\rho_0(\vecpar)$ under the newly determined design $\mat{x}_t$. We found that a batch-wise strategy of sampling variations can accelerate convergence. This strategy groups multiple variation samples with the newly acquired design sample, which helps the GP model to model the cost functional around the new design. Note that the grouping strategy is different from batch-wise Bayesian optimization, where a batch of design variables is determined from the acquisition function. %
A batch-wise version in solving $\mat{x}$ could potentially improve the DRBO efficiency and be of independent interest. 

Given the stop criteria of selecting the optimal $\mat{x}$ based on the posterior distribution, an augmented sampling set improves the estimation of this distribution, which in turn assists the algorithm in identifying the optimal design. The sample set can be augmented either by grouping more variation samples or by conducting more rounds of iterations. Further investigation is needed to achieve a good balance between these two strategies in terms of computational efficiency.

\section{Numerical Results}
\label{sec:numerical_results}
\begin{figure}[t]
    \centering
    \includegraphics[width = 3.3in]{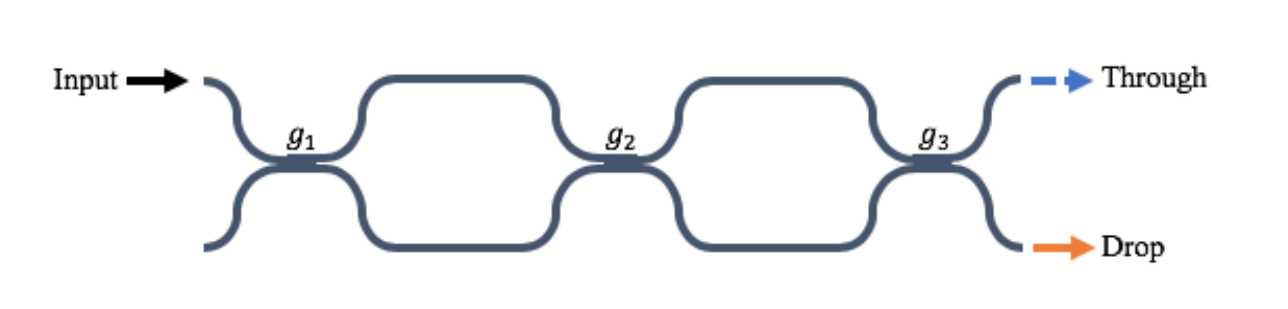}
    \caption{The schematic of a third-order Mach-Zehnder interferometer.}
    \label{fig:benchmark_MZI}
\end{figure}

We apply the proposed method to optimize a photonic IC and an electronic IC under various variation shifts.

\textbf{Baseline}: We compare the proposed DRBO algorithm with a standard LCB method~\cite{srinivas2009gaussian}, which serves as our baseline. LCB method assumes that the process variation follows a fixed distribution $\rho_0(\vect{\xi})$, whose acquisition function~\eqref{eq:chi_square_acq} degenerates to $ A(\mat{x}) = \frac{1}{L} \sum_{l=1}^L[\mu(\mat{x},\vect{\xi}^l)-\sqrt{\beta}\sigma(\mat{x},\vect{\xi}^l)]$. 
Therefore, it does not consider the variation shifts and thus operates as a standard Bayesian optimization method. Sharing a similar computational cost, it can also be viewed as a DRBO with $\varepsilon=0$. Specifically, this is exactly what existing Bayesian optimization solvers do in statistical circuit design optimization.

\textbf{Algorithm evaluation.} 
In this comparison, both DRBO and LCB are implemented with the same nominal variation distribution $\rho_0(\vecpar)$ and with the same initial setup to search optimal designs. To evaluate the robustness of these designs against distribution shifts, we then test the resulting designs under various unforeseen variation distributions. %

We have introduced three kinds of variation shifts in Section~\ref{sec:variation_shift}. 
For the photonics benchmark, we assume that the variation shifts arise due to {\it time shift}. In the electronic IC benchmark, we assume that the variation shifts arise from {\it poor data quality} and {\it model misfit}. In realistic applications, it is highly possible that all three scenarios would coexist.

\subsection{Photonics IC: Mach-Zehnder Interferometer}

\begin{figure}[t]
    \centering
    \vspace{-15pt}
    \includegraphics[width = 3.3in]{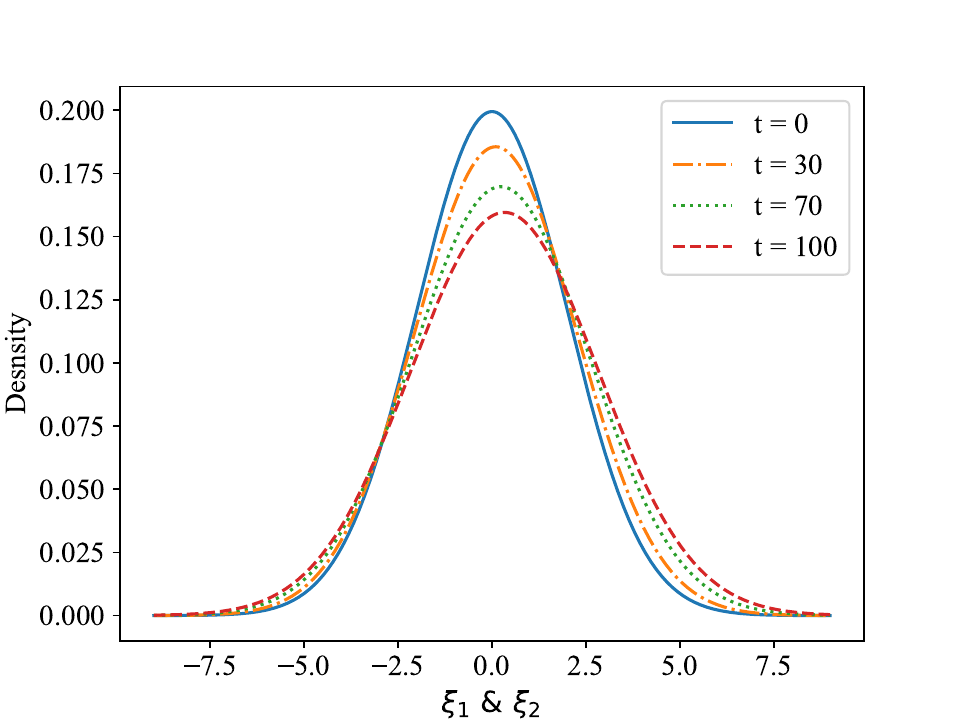}
    \caption{Examples of PDFs for the shifted (testing) variation distributions at different time steps, changing from the nominal distribution at $t = 0$ to a highly shifted one at $t = 100$.}
    \label{fig:test_sample}
\end{figure}

\textbf{Setup.} We consider a third-order Mach-Zehnder interferometer (MZI) which consists of three port coupling and two arms, as shown in Fig.~\ref{fig:benchmark_MZI}. 
By fixing the first gap parameter, we design the other two gap parameters $\mat{x} = [g_1, g_2] \in {[100,300]}^2 ~\text{nm}$ of the two arms under variations $\vect{\xi} \in \R^2$. 
We aim to maximize the expected 3-dB bandwidth (BW, in GHz) with risk constraints on the crosstalk (XT, in dB) and the attenuation ($\alpha$, in dB) of the peak transmission. The design objective is defined as ${f(\mat{x},\vect{\xi}) = -\text{BW}(\mat{x},\vect{\xi}) + \lambda I(\mat{x},\vect{\xi})}$ with 
\begin{equation}\label{eq:MZI_constraints}
I(\mat{x},\vect{\xi}) = 
    \begin{cases}
    0, & \text{XT}(\mat{x},\vect{\xi})\le \text{XT}_0, \alpha(\mat{x},\vecpar)\le \alpha_0\\
    1, & \text{otherwise}.
    \end{cases}   
\end{equation}

Here, we aim to evaluate the design robustness under the time shift of the variation distribution. Without loss of generalization, we assume the PDFs of the two variations follow the same time-dependent Gaussian distribution 
\begin{align*}
    &\rho(\xi_1(t)) = \rho(\xi_2(t)) \sim \mathcal{N} (\mu(t), \sigma(t)) ~\text{with}\\
    &\mu(t) = t/300, ~\sigma(t) = 2 + 0.005t.
\end{align*}
The PDFs of the resulting time-shift variations are shown in Fig.~\ref{fig:test_sample}. Specifically, the nominal distribution $\rho_0$ corresponds to the variation PDF at $t=0$. 

\begin{figure*}[t]
    \centering
    \includegraphics[width = 1\linewidth]{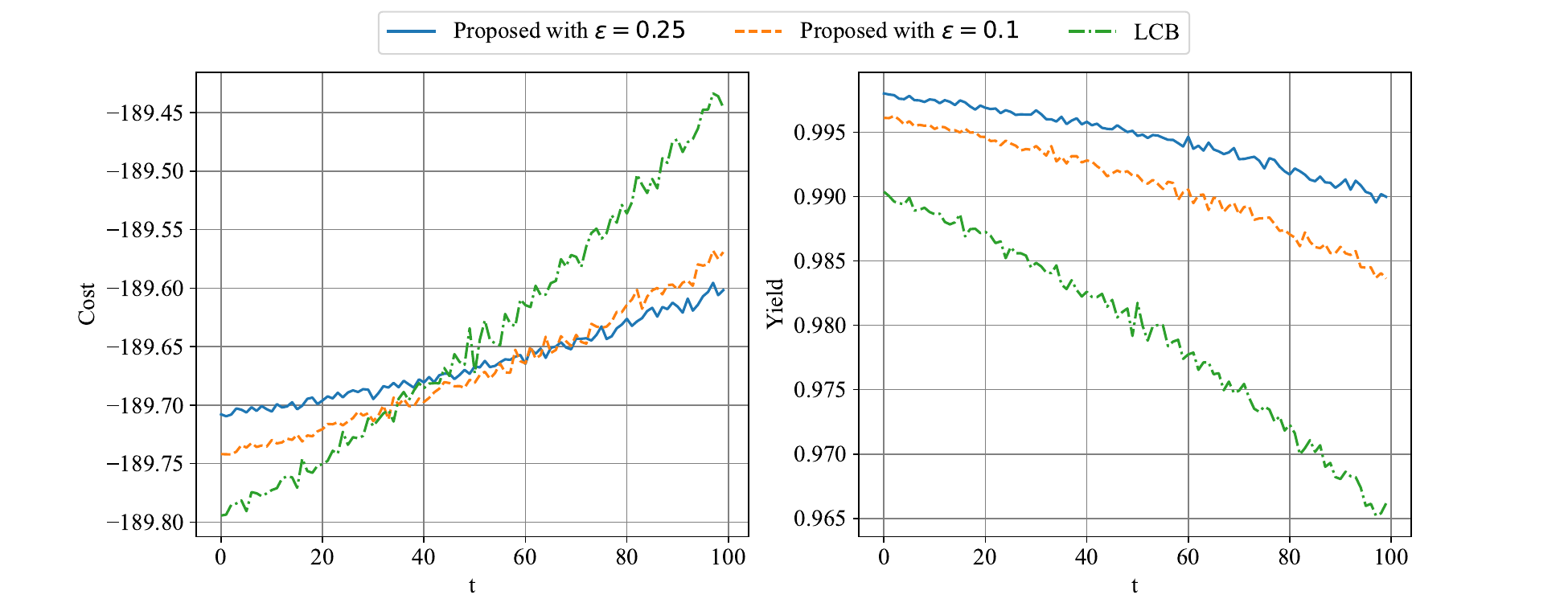}
    \caption{The optimized cost function and yield for the MZI under variation time shift. Regarding the cost function $f(\mat{x}, \vect{\xi})$, the LCB outperforms the DRBO ones when $t$ is relatively small. However, as $t$ increases, the DRBO methods begin to achieve better performance. The DRBO method with a larger $\varepsilon$ consistently achieves the highest yield, because it is more conservative in considering the worst case of variation shifts.}
    \label{fig:MZI_results}
\end{figure*}

For this benchmark, which involves two-dimensional design and variation variables, we begin by randomly selecting an initial sample set of $200$ pairs of $(\mat{x}, \vect{\xi})$. Here, $\mat{x}$ is uniformly sampled from the design variable range, and $\vect{\xi}$ is sampled from the nominal distribution. We conduct experiments using five different sets of initial samples for all algorithms, reporting the mean and standard deviation of the results.
To accelerate the learning process, after determining the next design in an iteration, we use the grouping strategy to pair this design with a batch of $5$ variation samples from the nominal distribution to augment the sample set. For all algorithms, we use $L=300$ samples in evaluating the acquisition function and set the maximum iteration count $T=150$ as the stopping criterion.

\textbf{Results.} 
We report the cost $f(\mat{x}, \vect{\xi})$ and yield of each iteration under shifted variations in Fig.~\ref{fig:MZI_results}. In terms of evaluation, a lower cost function is more desirable, while a higher yield is better. Specifically, we compare the performance among the LCB and the DRBO algorithm, using uncertainty set radius of $\varepsilon=0.25$ and $\varepsilon=0.1$. %
At each time step $t$, we obtain a design (not a query point) based on the model's posterior distribution and evaluate its performance and yield under the shifted variation distribution. 
The results at specific time points are detailed in Table~\ref{tab:MZI_design}.
Initially, at $t = 0$, the LCB outperforms the DRBO since the testing (true) distribution is exactly the nominal one. However, as $t$ increases, and the testing distribution begins to diverge from the nominal one, LCB begins to degrade, eventually becoming worse than DRBO after $t>30$. 
In contrast, the DRBO algorithms demonstrate increased robustness to variation shifts over time. Specifically, when the shift is relatively limited (e.g., when $t < 60$), the DRBO with $\varepsilon = 0.1$ performs better than the one with $\varepsilon = 0.25$ since a larger uncertainty set leads to a more conservative design optimization. However, when the shift becomes more pronounced, a larger $\varepsilon$ shows beneficial.

\begin{table*}[t]
\centering
\caption{Performance of the designs in the MZI benchmark tested under variations at different $t$.}
\label{tab:MZI_design}
\begin{adjustbox}{width=4in}
\begin{tabular}{ccccc}
\toprule
Time step  & Method & Cost function & Yield (\%)\\
\midrule 
\multirow{3}{*}{t = 0 (Nominal)}  
& {Proposed ($\varepsilon=0.25$)}  &{-189.69$\pm$0.099}  & 99.9$\pm$0.11\\
& {Proposed ($\varepsilon=0.10$)}  &{-189.74$\pm$0.085}  & 99.6$\pm$0.36\\
& \textbf{LCB~\cite{srinivas2009gaussian}} & \textbf{-189.80$\pm$0.015}  & 99.1$\pm$0.41 \\
\hline
\multirow{3}{*}{t = 30}  
& {Proposed ($\varepsilon=0.25$)}  &{-189.68$\pm0.087$}  & 99.7$\pm$0.21\\
& \textbf{{Proposed ($\varepsilon=0.10$)}}  &\textbf{-189.71$\pm$0.049}  & 99.4$\pm$0.64\\
& LCB~\cite{srinivas2009gaussian}  &-189.71$\pm$0.044  & 98.6$\pm$0.61\\
\hline
\multirow{3}{*}{t = 70}  
& \textbf{Proposed ($\varepsilon=0.25$)}  &\textbf{-189.65$\pm$0.069}  & 99.5$\pm0.35$\\
& Proposed ($\varepsilon=0.10$)  &-189.65$\pm0.037$  & 98.9$\pm$0.96\\
& LCB~\cite{srinivas2009gaussian} & -189.57$\pm$0.094  & 97.5$\pm$0.92 \\
\hline
\multirow{3}{*}{t = 100}  
& \textbf{Proposed ($\varepsilon=0.25$)}  &\textbf{-189.61$\pm$0.059}  & 99.3$\pm$0.57\\
& {Proposed ($\varepsilon=0.10$)}  &{-189.57$\pm$0.093}  & 98.3$\pm$1.45\\
& LCB~\cite{srinivas2009gaussian} & -189.45$\pm$0.124  & 96.6$\pm$1.17 \\
\bottomrule
\end{tabular}
\end{adjustbox}
\end{table*}

\textbf{Parameter Analysis.} 
We analyze the sample size $L$ used in evaluating the acquisition function~\eqref{eq:chi_square_acq}. For each setting, we present the mean and standard deviation of five repeated experiments with different initial samples. Given that we use a finite number of samples to estimate the nominal distribution, more samples from the distribution lead to a more accurate estimation, and consequently, more robust optimization and design. As shown in Table~\ref{tab:Lsize_MZI}, with the increase of $L$, the design performance generally improves, evidenced by a lower mean and smaller standard deviation in the cost function $f(\mat{x}, \vect{\xi})$.

However, increasing the number of samples $L$ in estimating the mean and standard deviation inherently takes more computational time, leading to a trade-off between computational time and robustness. In the benchmark presented, the problem size is relatively small, so the inference time in a GP model can be almost ignored. However, for higher-dimensional problems, the computational cost could become a significant factor. There are also many potential accelerations towards the DRBO solvers, such as parallelizing the optimization of the acquisition function by starting from different initial points, using the strategy to augment the sample set as previously discussed, and so forth.
\begin{table}[t]
\centering
\caption{Resulting cost function of DRBO with $\varepsilon=0.25$ when using different $L$ in optimizing the MZI.}
\label{tab:Lsize_MZI}
\begin{adjustbox}{width=3.6in}
\begin{tabular}{ccccc}
\toprule
$L$ & $t = 0$ & t = 30  &t = 70 & t = 100 \\
\midrule 
100 &-189.68$\pm$0.104 & -189.65$\pm$0.093 &-189.58$\pm$0.135    &-189.52$\pm$0.161\\
\hline
200  &-189.70$\pm$0.058 & -189.67$\pm$0.039 &-189.62$\pm$0.065    &-189.57$\pm$0.125\\
\hline
300 &-189.69$\pm$0.094 & -189.68$\pm$0.087 &-189.65$\pm$0.068    &-189.61$\pm$0.056\\
\hline
400 &-189.70$\pm$0.085 & -189.69$\pm$0.070 &-189.66$\pm$0.051    &-189.62$\pm$0.026\\
\hline
500 &\textbf{-189.75$\pm$0.038} & \textbf{-189.72$\pm$0.021} &\textbf{-189.68$\pm$0.016}   &\textbf{-189.62$\pm$0.036} \\
\bottomrule
\end{tabular}
\end{adjustbox}
\end{table}

\subsection{Electronics IC: Two-stage Amplifier}
\begin{figure}[t]
    \centering
    \includegraphics[width = 3.3in]{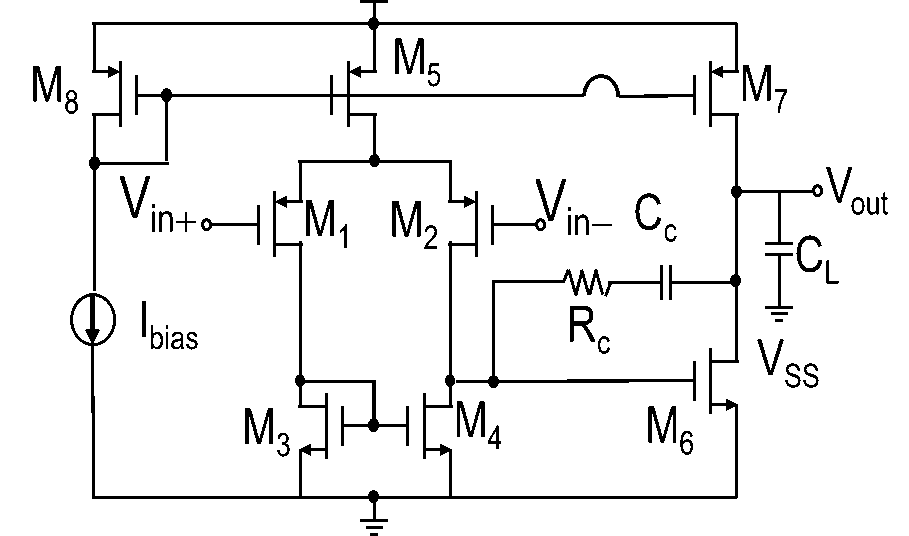}
    \caption{The schematic of a two-stage amplifier.}
    \label{fig:benchmark_amplifier}
\end{figure}

\textbf{Setup.} We consider a two-stage operational amplifier (Op-Amp) whose schematic is shown in Fig.~\ref{fig:benchmark_amplifier}. Let $W_i$ denote the width (in $\mu m$) of the transistor $M_i$. %
To reduce the systematic offset voltage, we enforce $W_1 = W_2$, $W_3 = W_4$, and set a ratio $R = W_7/W_5 = W_6/2 W_3$. We consider $ \mat{x} = [W_1, W_6, W_7, R]$ as the design variables for optimization. However, due to the process variations in manufacturing, the enforced equalities may not hold true. Consequently, we consider $\vect{\xi} = [\epsilon_2, \epsilon_4]$ as the process variations, where $W_2 =  (1 + \epsilon_2) W_1 $ and $W_4 = (1 + \epsilon_4) W_3 $. These variations have been found to significantly impact the circuit's performance~\cite{zheng2020}.
Here we aim to minimize the power (in mw) while having risk constraints on the gain (in dB), unity gain frequency (UGF, in MHZ), and phase margin (PM, in degree). The design objective is defined as: ${f(\mat{x},\vect{\xi}) = -\text{power}(\mat{x},\vecpar) + \lambda I(\mat{x},\vect{\xi})}$ where $I(\mat{x},\vect{\xi}) = 0$ when $\text{gain}(\mat{x},\vect{\xi})\geq 30$, $\text{UGF}(\mat{x},\vect{\xi})\geq 120$ and $\text{PM}(\mat{x},\vect{\xi})\geq 60$ %
are satisfied, otherwise $I(\mat{x},\vect{\xi}) = 1$. 

We assume that the true parameter distribution $\rho(\vecpar)$ follows a (multivariate) Gaussian mixture model (GMM), which is a weighted sum of multiple (multivariate) Gaussian distributions:
\begin{align*}
   \rho(\vect{\xi}) = \sum_{i=1}^n \alpha_i \mathcal{N}(\vect{\mu}_i, \vect{\Sigma}_i), \quad \text{with} \quad \sum_{i=1}^n \alpha_i = 1.
\end{align*}
In this study, we set $n = 2$. The variables $\boldsymbol{\mu}_1$ and $\boldsymbol{\mu}_2$ denote the means of the two Gaussian distributions, $\boldsymbol{\Sigma}_1$ and $\boldsymbol{\Sigma}_2$ denote their covariance matrices, and $\alpha_1$ and $\alpha_2$ represent their respective mixing weights. Specifically, we set $\boldsymbol{\mu}_1(\vect{\xi}) = -\boldsymbol{\mu}_2(\vect{\xi}) = [0.008, 0.008]$ and $\boldsymbol{\Sigma}_1(\vect{\xi}) = \boldsymbol{\Sigma}_2(\vect{\xi}) = 10^{-2} \begin{bmatrix} 0.8 & 0.1 \\ 0.1 & 0.8 \end{bmatrix}$. 

\begin{figure}[t]
    \centering
    \includegraphics[width = 3.4in]{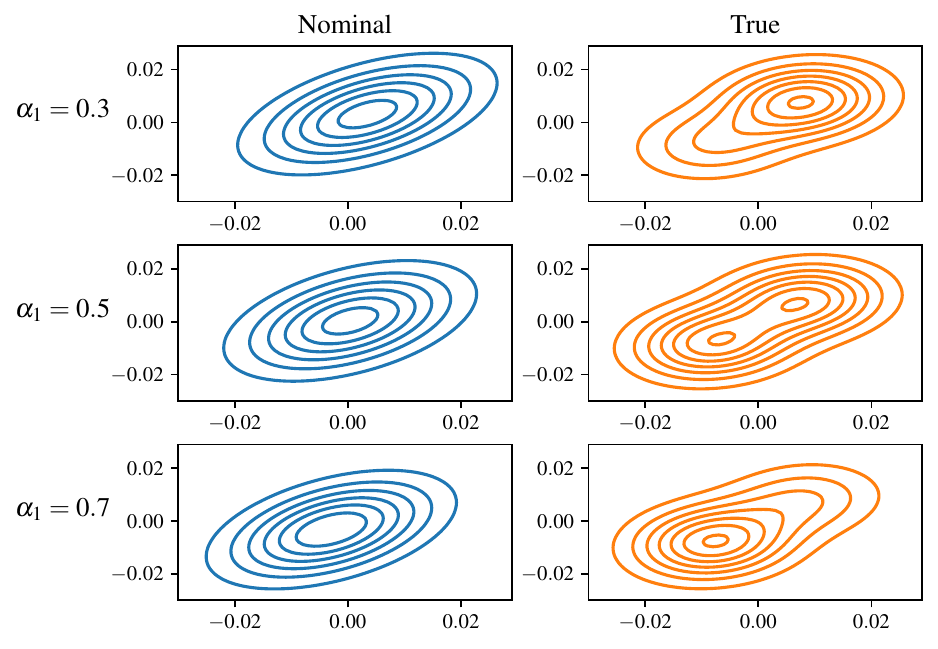}
    \caption{PDFs of the nominal Gaussian distribution $\rho_0(\vecpar)$ and the true Gaussian mixture distribution $\rho(\vecpar)$. We consider three cases where the true variations are a mixture of two Gaussian distributions with $\alpha_1 = 0.3, 0.5,$ and $0.7$.}
    \label{fig:distribution_misfit}
\end{figure}

To simulate variation shifts due to model misfit and poor data quality, we draw $500$ noisy samples from the true GMM $\rho(\vect{\xi})$ and fit a single Gaussian distribution to these samples. This fitted Gaussian serves as our nominal distribution $\rho_0(\vecpar)$. The poor data quality is represented by adding noise to the samples from the true distribution: $\vect{\xi}_{\text{noisy}} = \vect{\xi} + \vect{\eta}$, where $\vect{\eta} \sim \mathcal{N}(\mat{0},10^{-4} \mat{I})$. In Fig.~\ref{fig:distribution_misfit}, we can clearly see the divergence between $\rho_0(\vect{\xi})$ and $\rho(\vect{\xi})$ under different mixing weights $\alpha_1$.

Similar to the previous MZI example, for this benchmark involving four design variables and two process variation variables, we randomly select $500$ pairs of $(\mat{x}, \vect{\xi})$ as the initial samples. During the Bayesian iterations, we group each query point with $10$ $\vect{\xi}$. We repeat the experiments using five different sets of initial samples for all algorithms, subsequently reporting the mean and standard deviation of the results. We set the maximum iteration $T=160$, and use $L=300$ for~\eqref{eq:chi_square_acq}. To further stabilize the design solution in this benchmark, we take the final optimal design as the output design with the lowest cost function from the last five Bayesian optimization iterations.

\begin{table}[t]
\centering
\caption{Performance of the designs in the two-stage amplifier benchmark tested under different distributions.}
\vspace{-1mm}
\label{tab:two_stage_amplifier}
\begin{adjustbox}{width=3.3in}
\begin{tabular}{cccccc}
\toprule
$\alpha_1$  & Testing $\rho(\vect{\xi})$ & Method & Cost function  & Yield (\%)\\
\midrule 
\multirow{4}{*}{0.3}  
& \multirow{2}{*}{Nominal} & Proposed  &0.2088$\pm$0.0037  &  93.32$\pm$1.36 \\
& & LCB~\cite{srinivas2009gaussian} &0.2124$\pm$0.0034   & 91.73$\pm$2.02 \\ 
\addlinespace[0.2ex]
\cline{2-5}
\addlinespace[0.5ex]
& \multirow{2}{*}{True} & Proposed  &\textbf{0.2195$\pm$0.0043}   &  91.63$\pm$2.02 \\
& & LCB~\cite{srinivas2009gaussian} &  0.2282$\pm$0.0159  &  86.63$\pm$4.01 \\
\addlinespace[0.2ex]
\hline
\addlinespace[0.5ex]
\multirow{4}{*}{0.5}  
& \multirow{2}{*}{Nominal} & Proposed & 0.2142$\pm$0.0035  & 91.91$\pm$2.91 \\
& & LCB~\cite{srinivas2009gaussian} &0.2182$\pm$0.0047   & 90.70$\pm$2.20 \\
\addlinespace[0.2ex]
\cline{2-5}
\addlinespace[0.5ex]
& \multirow{2}{*}{True} & Proposed & \textbf{0.2191$\pm$0.0039}  & 91.36$\pm$3.04 \\
& & LCB~\cite{srinivas2009gaussian} &0.2290$\pm$0.0115   & 87.78$\pm$1.47  \\
\addlinespace[0.2ex]
\hline
\addlinespace[0.5ex]
\multirow{4}{*}{0.7}  
& \multirow{2}{*}{Nominal} & Proposed & 0.2241$\pm$0.0067 &  88.09$\pm$0.98\\
& & LCB~\cite{srinivas2009gaussian} & 0.2256$\pm$0.0112  & 88.07$\pm$3.19 \\
\addlinespace[0.2ex]
\cline{2-5}
\addlinespace[0.5ex]
& \multirow{2}{*}{True} & Proposed & \textbf{0.2231$\pm$0.0104}   &  88.13$\pm$1.76\\
& & LCB~\cite{srinivas2009gaussian} & 0.2287$\pm$0.0144  & 87.11$\pm$3.72 \\
\bottomrule
\end{tabular}
\end{adjustbox}
\vspace{-3mm}
\end{table}

\begin{table}[t]
\centering
\caption{Parameter analysis on the radius $\varepsilon$ of the uncertainty ball under true distribution with $\alpha_1 = 0.3$.}
\vspace{-1mm}
\label{tab:hspice_errorsize}
\begin{adjustbox}{width=2.5in}
\begin{tabular}{cccc}
\toprule
Radius $\varepsilon$ & Cost function  & Yield (\%)\\
\midrule 
0.15  &0.2287$\pm$0.0144    &  92.71$\pm$0.94 \\
0.10  &\textbf{0.2181$\pm$0.0044} &  92.54$\pm$1.61 \\
0.05  &0.2195$\pm$0.0043   &  91.63$\pm$2.02    \\
0.01  &0.2257$\pm$ 0.0057    &  88.71$\pm$1.17  \\
0 (LCB)  &  0.2282$\pm$0.0159  &  86.63$\pm$4.01   \\
\bottomrule
\end{tabular}
\end{adjustbox}
\vspace{-3mm}
\end{table}

\textbf{Results.} 
We evaluate the design performance of both LCB and DRBO with $\varepsilon = 0.05$ under three specific cases of variation shifts, namely, where the true distributions have different mixing weights $\alpha_1= 0.3, 0.5,$ and $0.7$. Additionally, we also test the design performance under the nominal distributions as a reference. For each design, its cost function and yield are estimated based on $5\times 10^3$ simulated samples. The resulting design performance and comparisons are comprehensively outlined in Table~\ref{tab:two_stage_amplifier}.  
When we evaluate the designs under the nominal distributions, both the DRBO and LCB methods achieve comparable results, with a similar cost function and yield. This shows the effective optimization of two methods in the absence of variation shift. 
However, when we evaluate the designs under the true distributions, the DRBO solution demonstrates significantly better robustness under the shifts. This is evidenced by the lower cost function and higher yield associated with the DRBO solution. The observations are consistent across all three true distribution cases.

\textbf{Parameter Analysis.} 
We analyze the impact of the radius $\varepsilon$ of the uncertainty ball under the true variations where $\alpha_1 = 0.3$. The radius $\varepsilon$ represents our estimation of the potential variation shifts. In this testing case, the true $\chi^2$-divergence between the true PDF and nominal PDF is approximately $0.1$. Table~\ref{tab:hspice_errorsize} depicts the results obtained from DRBO with different values of $\varepsilon$. As we can see, with the increase in $\varepsilon$, the obtained cost function first decreases and then increases. The optimal cost function is achieved when the value of $\varepsilon$ is closest to the actual divergence. It indicates that in real-world applications, users should select the value of $\varepsilon$ based on their estimation of variation shifts. While we observe a consistent improvement in yield as $\varepsilon$ increases in this case, the monotonic relationship between the radius $\varepsilon$ and the yield is not strictly guaranteed in our yield-aware optimization problem. To prioritize yield optimization, we can simply assign a large value for the penalty factor $\lambda$. Meanwhile, other variants of distributionally robust yield optimization formulations deserve further investigation.

\section{Conclusion}
In practical IC design, we only have limited and inexact statistical knowledge of process variations. Therefore the actual PDF of process variations is quite uncertain, and it often differs from the given statistical model. In this paper, we outlined three possible scenarios of variation shift - poor data quality, model misfit, and time shift - which are likely to occur concurrently in real-world applications. To find a design that is robust against such variation shifts, we formulated a novel shift-aware circuit optimization problem as a distributionally robust optimization problem. This formulation optimizes the circuit design when the actual variation distribution is different from the nominal one. With the appropriate modeling of the uncertainty set for variation distributions, we effectively and efficiently solve the proposed shift-aware circuit optimization via a distributionally robust Bayesian optimization (DRBO) approach. We validate the proposed method using two IC benchmarks - a photonic IC and an electronic IC. Our approach successfully solves out designs that demonstrate greater robustness against variation shifts, whereas traditional stochastic optimization methods significantly underperform under the variation shifts.

Possible extensions include but are not limited to: (i) determining the radius of the uncertainty ball, (ii) techniques to handle high-dimensional design variables and process variations in complicated circuits, and (iii) further acceleration of the distributionally robust Bayesian optimization algorithms.

\section*{Acknowledgement}
Z. He and Z. Zhang were supported by NSF CCF \#1846476.

\small{
\bibliographystyle{Bib/IEEEtran}
\bibliography{Bib/reference}

\begin{thebibliography}{10}
\providecommand{\url}[1]{#1}
\csname url@samestyle\endcsname
\providecommand{\newblock}{\relax}
\providecommand{\bibinfo}[2]{#2}
\providecommand{\BIBentrySTDinterwordspacing}{\spaceskip=0pt\relax}
\providecommand{\BIBentryALTinterwordstretchfactor}{4}
\providecommand{\BIBentryALTinterwordspacing}{\spaceskip=\fontdimen2\font plus
\BIBentryALTinterwordstretchfactor\fontdimen3\font minus
  \fontdimen4\font\relax}
\providecommand{\BIBforeignlanguage}[2]{{%
\expandafter\ifx\csname l@#1\endcsname\relax
\typeout{** WARNING: IEEEtran.bst: No hyphenation pattern has been}%
\typeout{** loaded for the language `#1'. Using the pattern for}%
\typeout{** the default language instead.}%
\else
\language=\csname l@#1\endcsname
\fi
#2}}
\providecommand{\BIBdecl}{\relax}
\BIBdecl

\bibitem{singhee2010quasi}
A.~Singhee and R.~A. Rutenbar, ``Why {quasi-Monte Carlo} is better than {Monte
  Carlo} or {Latin} hypercube sampling for statistical circuit analysis,''
  \emph{IEEE Trans. Comput.-Aided Design Integr. Circuits Syst.}, vol.~29,
  no.~11, pp. 1763--1776, 2010.

\bibitem{stievano2011parameters}
I.~S. Stievano, P.~Manfredi, and F.~G. Canavero, ``Parameters variability
  effects on multiconductor interconnects via {Hermite} polynomial chaos,''
  \emph{IEEE Trans. CPMT}, vol.~1, no.~8, pp. 1234--1239, 2011.

\bibitem{zhang2013stochastic}
Z.~Zhang, T.~A. El-Moselhy, I.~M. Elfadel, and L.~Daniel, ``Stochastic testing
  method for transistor-level uncertainty quantification based on generalized
  polynomial chaos,'' \emph{IEEE Trans. CAD Integr. Circuits Syst.}, vol.~32,
  no.~10, pp. 1533--1545, 2013.

\bibitem{zhang2014enabling}
Z.~Zhang, X.~Yang, I.~V. Oseledets, G.~E. Karniadakis, and L.~Daniel,
  ``Enabling high-dimensional hierarchical uncertainty quantification by
  {ANOVA} and tensor-train decomposition,'' \emph{IEEE Trans. CAD Integr.
  Circuits Syst.}, vol.~34, no.~1, pp. 63--76, 2014.

\bibitem{he2021high}
Z.~He and Z.~Zhang, ``High-dimensional uncertainty quantification via tensor
  regression with rank determination and adaptive sampling,'' \emph{IEEE Trans.
  CMPT}, vol.~11, no.~9, pp. 1317--1328, 2021.

\bibitem{QBPV}
I.~Stevanovic and C.~C. McAndrew, ``Quadratic backward propagation of variance
  for nonlinear statistical circuit modeling,'' \emph{IEEE Trans. CAD of
  Integrated Circuits and Systems}, vol.~28, no.~9, pp. 1428--1432, 2009.

\bibitem{ReviewSM}
X.~Li, J.~Le, and L.~T. Pileggi, ``Statistical performance modeling and
  optimization,'' \emph{Found. Trends Electron. Des. Autom.}, vol.~1, no.~4, p.
  331–480, apr 2006.

\bibitem{zhang2016big}
Z.~Zhang, T.-W. Weng, and L.~Daniel, ``Big-data tensor recovery for
  high-dimensional uncertainty quantification of process variations,''
  \emph{IEEE Trans. CMPT}, vol.~7, no.~5, pp. 687--697, 2017.

\bibitem{srivastava2005statistical}
A.~Srivastava, D.~Sylvester, and D.~Blaauw, \emph{Statistical analysis and
  optimization for {VLSI}: Timing and power}.\hskip 1em plus 0.5em minus
  0.4em\relax Springer, 2005, vol.~59.

\bibitem{liu2011efficient}
B.~Liu, F.~V. Fern{\'a}ndez, and G.~G. Gielen, ``Efficient and accurate
  statistical analog yield optimization and variation-aware circuit sizing
  based on computational intelligence techniques,'' \emph{IEEE Trans. CAD of
  Integrated Circuits and Systems}, vol.~30, no.~6, pp. 793--805, 2011.

\bibitem{wang2017efficient}
M.~Wang, W.~Lv, F.~Yang, C.~Yan, W.~Cai, D.~Zhou, and X.~Zeng, ``Efficient
  yield optimization for analog and {SRAM} circuits via {G}aussian process
  regression and adaptive yield estimation,'' \emph{IEEE Trans. CAD of
  Integrated Circuits and Systems}, vol.~37, no.~10, pp. 1929--1942, 2017.

\bibitem{cui2020chance}
C.~Cui, K.~Liu, and Z.~Zhang, ``Chance-constrained and yield-aware optimization
  of photonic {ICs} with non-{Gaussian} correlated process variations,''
  \emph{IEEE Transactions on Computer-Aided Design of Integrated Circuits and
  Systems}, vol.~39, no.~12, pp. 4958--4970, 2020.

\bibitem{he2021pobo}
Z.~He and Z.~Zhang, ``{PoBO}: A polynomial bounding method for
  chance-constrained yield-aware optimization of photonic ics,'' \emph{IEEE
  Trans. CAD Integr. Circuits Syst.}, vol.~41, no.~11, pp. 4915--4926, 2022.

\bibitem{zhang2020bayesian}
S.~Zhang, F.~Yang, D.~Zhou, and X.~Zeng, ``Bayesian methods for the yield
  optimization of analog and {SRAM} circuits,'' in \emph{Asia and South Pacific
  Design Automation Conf.}, 2020, pp. 440--445.

\bibitem{antreich1994circuit}
K.~J. Antreich, H.~E. Graeb, and C.~U. Wieser, ``Circuit analysis and
  optimization driven by worst-case distances,'' \emph{IEEE Trans.
  Comput.-Aided Design Integr. Circuits Syst.}, vol.~13, no.~1, pp. 57--71,
  1994.

\bibitem{dharchoudhury1995worst}
A.~Dharchoudhury and S.-M. Kang, ``Worst-case analysis and optimization of
  {VLSI} circuit performances,'' \emph{IEEE Trans. CAD Integr. Circuits Syst.},
  vol.~14, no.~4, pp. 481--492, 1995.

\bibitem{kirschner2020distributionally}
J.~Kirschner, I.~Bogunovic, S.~Jegelka, and A.~Krause, ``Distributionally
  robust {Bayesian} optimization,'' in \emph{International Conference on
  Artificial Intelligence and Statistics}.\hskip 1em plus 0.5em minus
  0.4em\relax PMLR, 2020, pp. 2174--2184.

\bibitem{nguyen2020distributionally}
T.~Nguyen, S.~Gupta, H.~Ha, S.~Rana, and S.~Venkatesh, ``Distributionally
  robust {Bayesian} quadrature optimization,'' in \emph{International
  Conference on Artificial Intelligence and Statistics}.\hskip 1em plus 0.5em
  minus 0.4em\relax PMLR, 2020, pp. 1921--1931.

\bibitem{husain2022distributionally}
H.~Husain, V.~Nguyen, and A.~v.~d. Hengel, ``Distributionally robust {Bayesian}
  optimization with $\phi$-divergences,'' \emph{arXiv preprint
  arXiv:2203.02128}, 2022.

\bibitem{stine1997analysis}
B.~E. Stine, D.~S. Boning, and J.~E. Chung, ``Analysis and decomposition of
  spatial variation in integrated circuit processes and devices,'' \emph{IEEE
  Trans. Semiconductor Manufacturing}, vol.~10, no.~1, pp. 24--41, 1997.

\bibitem{gmm15}
A.~Lange, C.~Sohrmann, R.~Jancke, J.~Haase, B.~Cheng, A.~Asenov, and
  U.~Schlichtmann, ``Multivariate modeling of variability supporting
  non-gaussian and correlated parameters,'' \emph{IEEE Trans. CAD of Integrated
  Circuits and Systems}, vol.~35, no.~2, pp. 197--210, 2016.

\bibitem{yu2013statistical}
L.~Yu, L.~Wei, D.~Antoniadis, I.~Elfadel, and D.~Boning, ``Statistical modeling
  with the virtual source mosfet model,'' in \emph{Design, Automation \& Test
  in Europe Conference \& Exhibition (DATE)}, 2013, pp. 1454--1457.

\bibitem{raj2011nanoscale}
B.~Raj, A.~Saxena, and S.~Dasgupta, ``Nanoscale {FinFET} based {SRAM} cell
  design: Analysis of performance metric, process variation, underlapped
  {FinFET}, and temperature effect,'' \emph{IEEE Circuits and Systems
  Magazine}, vol.~11, no.~3, pp. 38--50, 2011.

\bibitem{reddy2004impact}
V.~Reddy, J.~Carulli, A.~Krishnan, W.~Bosch, and B.~Burgess, ``Impact of
  negative bias temperature instability on product parametric drift,'' in
  \emph{International Conference on Test}, 2004, pp. 148--155.

\bibitem{han2011}
S.~Han, J.~Choung, B.-S. Kim, B.~H. Lee, H.~Choi, and J.~Kim, ``Statistical
  aging analysis with process variation consideration,'' in \emph{International
  Conference on Computer-Aided Design}, 2011, pp. 412--419.

\bibitem{wang2007efficient}
W.~Wang, Z.~Wei, S.~Yang, and Y.~Cao, ``An efficient method to identify
  critical gates under circuit aging,'' in \emph{2007 IEEE/ACM International
  Conference on Computer-Aided Design}.\hskip 1em plus 0.5em minus 0.4em\relax
  IEEE, 2007, pp. 735--740.

\bibitem{rahimian2019distributionally}
H.~Rahimian and S.~Mehrotra, ``Distributionally robust optimization: A
  review,'' \emph{arXiv preprint arXiv:1908.05659}, 2019.

\bibitem{f_diver_1966}
S.~M. Ali and S.~D. Silvey, ``A general class of coefficients of divergence of
  one distribution from another,'' \emph{Journal of the Royal Statistical
  Society: Series B (Methodological)}, vol.~28, no.~1, pp. 131--142, 1966.

\bibitem{williams2006gaussian}
C.~K. Williams and C.~E. Rasmussen, \emph{Gaussian processes for machine
  learning}.\hskip 1em plus 0.5em minus 0.4em\relax MIT press Cambridge, MA,
  2006, vol.~2, no.~3.

\bibitem{cakmak2020bayesian}
S.~Cakmak, R.~Astudillo~Marban, P.~Frazier, and E.~Zhou, ``Bayesian
  optimization of risk measures,'' \emph{Advances in Neural Information
  Processing Systems}, vol.~33, pp. 20\,130--20\,141, 2020.

\bibitem{frazier2018tutorial}
P.~I. Frazier, ``A tutorial on bayesian optimization,'' \emph{arXiv preprint
  arXiv:1807.02811}, 2018.

\bibitem{srinivas2009gaussian}
N.~Srinivas, A.~Krause, S.~Kakade, and M.~Seeger, ``Gaussian process
  optimization in the bandit setting: No regret and experimental design,'' in
  \emph{ICML}, 2010, p. 1015–1022.

\bibitem{zheng2020}
Z.~Gao and R.~Rohrer, ``Efficient non-monte-carlo yield estimation,''
  \emph{IEEE Transactions on Computer-Aided Design of Integrated Circuits and
  Systems}, vol.~41, no.~5, pp. 1222--1235, 2022.

\end{thebibliography}
}

\end{document}